\documentclass[prl,aps,amsmath,amssymb,twocolumn]{revtex4}
\usepackage{graphicx}
\begin{document}
\title{Oscillations of magnetoresistance in a clean hollow cylinder with fluctuating radius}
\author{A.S.Ioselevich}
\address{L.D.Landau Institute for Theoretical Physics, Moscow
119334, Russia,\\
Moscow Institute of Physics and Technology, Moscow 141700,
Russia.}
\date{\today}

\begin{abstract}
We consider magnetic oscillations of resistivity of a clean (mean free path $l\gg R$) hollow cylinder with fluctuating (with an amplitude of fluctuations $\Delta R\ll\overline{R}$) radius $R$, threaded by magnetic flux $\Phi$. We demonstrate, that for weak  fluctuations ($\Delta R\ll p_F^{-1}$) the oscillations have a standard period $2\Phi_0$, characteristic for oscillations in a clean system, while for $\Delta R\gg p_F^{-1}$ they become $\Phi_0$-periodic, which was expected only for dirty systems with $l\ll R$. The work is motivated by observation of predominantly $\Phi_0$-periodic magnetic oscillations in very clean Bismuth wires.
\end{abstract}

\maketitle

{\bf Introduction}. Oscillations of resistivity $\rho$ of  a hollow conducting cylinder with change of magnetic field $H$, parallel to the axis of the cylinder, is a well known phenomenon, manifesting the quantum nature of conducting electrons in metals. The basic reason for these oscillations is the interference of electrons traversing the cylinder along paths with different topological winding numbers. In the simplest case of clean cylinder (with the circumference $2\pi R\ll l$) it is actually a direct realization of the Aharonov-Bohm effect \cite{AB59}; the resistance  is a periodic function of the flux $\Phi=\pi R^2H$ with the period $2\Phi_0$, where $\Phi_0=\pi c\hbar/2e$ is the flux quantum (see, e.g., \cite{Brandt77,Brandt82}).    The general arguments (c.f. \cite{Imry}) show that the same periodicity should be observed for any thermodynamic or kinetic property of a cylinder -- for arbitrary relation between $2\pi R$ and $l$. In particular, for the resistivity
\begin{align}
\rho=\rho_0\left(1+\sum_{n=1}^\infty A_n\cos(\pi n\Phi/\Phi_0)\right)
	\label{gen-rho}
\end{align}
In dirty cylinders, where $2\pi R\gg l$ the odd-$n$ harmonics of the AB-oscillations are washed out due to strong variations in the length of different diffusive trajectories that lead to randomisation of the non-magnetic part of the phases of electronic wave-functions. In a seminal paper by Altshuler, Aronov and Spivak \cite{AAS81} it was shown, however, that the even-$n$ harmonics -- the oscillations, associated with a special sort of trajectories (the ones, containing  closed topologically non-trivial loops on the cylinder) survive the randomisation. These even-$n$ oscillations arise due to the weak localisation corrections and manifest the interference between the contributions of  paths, corresponding to the same trajectories, but with opposite directions in which the loop is traversed. In such an interference the randomised nonmagnetic part of the phase does not enter, while the magnetic part of the phase is doubled and, therefore, the AAS-oscillations occur with the period $\Phi_0$.   The AAS-oscillations were observed experimentally (see \cite{AASSS82,AS87}), although their amplitude is small:  $A_n^{(\rm AAS)}\sim (p_Fl)^{-1}$.

Thus, a simple correspondence seemed to be established: In clean cylinders with $2\pi R\ll l$ the AB-oscillations with the period $2\Phi_0$ should dominate, while dirty cylinders (with $2\pi R\gg l$) is the domain of the AAS-oscillations with the period $\Phi_0$. 

Recently, however,  the dominance of the $\Phi_0$-periodic (AAS-like) oscillations was experimentally observed \cite{Nikolaeva2008} in a manifestly clean Bismuth wires, where the AB-oscillations would be natural to expect. Note, that in this system the surface states are believed to be mainly responsible for the conductivity and the spin-orbit interaction is important for these states. In this situation the oscillations of resistivity were discussed in \cite{ZV2010}, where it was shown that {\it in the case of strong surface disorder} the odd harmonics in \eqref{gen-rho} should be suppressed, and the oscillations should be indeed $\Phi_0$-periodic, like in the AAS-case. However, the model, used in \cite{ZV2010} seems to be not applicable to the experiments  \cite{Nikolaeva2008}, since it implies strong disorder, which is apparently absent there.

Trying to explain thie unexpected experimental observation of \cite{Nikolaeva2008}, one should address the following question: 
 Suppose that there is some mechanism of suppression of odd-$n$ harmonics, then  how is it possible, that the same mechanism does not suppress also the mean free path $l$ and does not show up in the monotonic part of the resistivity? 
 
 On the other hand, under the condition $2\pi R\ll l$ the standard method of finding  even-$n$ AAS-oscillations, based on the diffusion approximation, does not apply, and even the qualitative physics underlying the  $\Phi_0$-periodic oscillations may be different from the weak localisation effect, considered in \cite{AAS81}.

In this Letter we propose a simple mechanism for this effect, related to weak and smooth fluctuations of the radius of the wire. Here we will discuss it only for the simplest case of weak short range impurities, that can be treated within the Born approximation; we also do not take into account the spin-orbital interaction. The latter effects will be considered in a separate publication, here we only mention, that they may, in some cases, considerably alter the amplitude of the even-$n$ harmonics, but they do not change the principal result: odd harmonics are suppressed much stronger than even ones.

We argue, that even relatively small fluctuations of the cylinders radius $R(z)$, measured at different cross-sections $z$, are able to exponentially  suppress the oscillations. Namely
\begin{align}
	A_n(\overline{Q},q)=a_n+\frac{2}{ \sqrt{ \pi n\overline{Q}}}e^{-2n^2q^2}\cos (2n\overline{Q}+\pi/4)\nonumber\\
	a_n=\frac{6}{\pi(n\overline{Q})^{3}}\left\{\begin{aligned}
	0,\qquad & \mbox{($n$ -- odd),}\\
	1,\qquad & \mbox{($n$ -- even),}
	\end{aligned}
	\right.
	\label{life-time6u}
\end{align}
where we have introduced the parameters
\begin{align}
\overline{Q}\equiv \pi \overline{R}p_F,\qquad q\equiv \pi \Delta Rp_F\ll \overline{Q},
	\label{param1}
\end{align}
$\overline{R}$ being the cylinders radius, averaged over the positions of the cross section, and  $\Delta R\equiv\left(\overline{(R-\overline{R})^2}\right)^{1/2}\ll \overline{R}$, being the relatively small variance of $R$.
The quasiclassical parameter $\overline{Q}$ we will assume to be large: $\overline{Q}\gg 1$, while $q$ may be either small, or large. 

We demonstrate that at $q\gg 1$ the fluctuations exponentially suppress all the amplitudes $A_n$; however, while the odd harmonics are suppressed to zero, the even ones saturate at some small (in parameter $\overline{Q}^{-1}$), but finite values $a_n$. Thus, at $q\to\infty$ the oscillations become $\Phi_0$-periodic, as in the dirty case.

 On the other hand, if the fluctuations of $R(z)$ are adiabatically smooth (their correlation radius $\xi\gg R$) then this kind of disorder leads to only exponentially small contribution to the scattering of electrons and, therefore, practically does not affect the mean free path $l$. 
 
 It should also be noted that our result \eqref{life-time6u}, in contrast with the result of \cite{AAS81}, is not proportional to small parameter $(p_Fl)^{-1}$: indeed it does not involve any weak localisation effects and does not depend on $l$ at all. Although the derivation of expression for the residual amplitude $a_n$ requires taking into account quantum corrections to the standard Drude formula, these corrections are not associated with the interference of different scattering events. It does not mean that the localisation corrections do not exist in our case -- they do, but for low concentration of impurities the corresponding corrections to $A_n$ are small.

{\bf A homogeneous tube. }We start with considering a hollow metallic cylinder with a fixed radius $R$ and length $L$, threaded by a magnetic flux $\Phi$ (the magnetic field is oriented along the axis of a cylinder $z$). The effective two-dimensional metal, constituting the tube, is characterized by an isotropic spectrum of electrons, $p_F$ and $v_F$ being, respectively,  the momentum and the velocity at the Fermi level $E_F$. There are rare impurities in the system -- with the two-dimensional concentration $n^{(2)}_{\rm imp}$, the corresponding mean free path being $l$. 
In this paper we restrict our consideration to the case of low concentration $n^{(2)}_{\rm imp}\ll R^{-2}$ and consider the semiclassical and clean case, when
\begin{align}
	p_F^{-1}\ll 2\pi R\ll \Delta z_{\rm imp},\quad l\ll L
	\label{main-cond1}
\end{align}
where $\Delta z_{\rm imp}\sim (2\pi Rn^{(2)}_{\rm imp})^{-1}\gg R$ is a characteristic separation between impurities.

In an absolutely clean system the eigenfunctions of the hamiltonian at the Fermi surface are
\begin{align}
	\psi_m(\phi,z)=\exp\{ik_mz+im\phi\},\\ k_m=\pm\sqrt{p_F^2-(m+\Phi/2\Phi_0)^2/R^2},
	\label{main-cond2}
\end{align}
where $m\in Z$ is the azymuthal quantum number.

Under the condition \eqref{main-cond1} the system can be treated as a multichannel quasi-one-dimensional one (the number of channels $N_{\rm ch} =(2\pi R)( p_F/2\pi)=p_FR\gg 1$), so that the localization length ${\cal L}_{\rm loc}= N_{\rm ch}l/2\gg l$   is large. We will assume that it is larger than the length of the wire:
\begin{align}
	l\ll L\ll {\cal L}_{\rm loc}=l(p_FR)/2
	\label{main-cond1a}
\end{align}
and the  specific one-dimensional localization effects are relatively small, and can be treated perturbatively (weak localisation). From this assumption it immediately follows, that, in the leading semiclassical approximation, the conductivity  of the wire $\sigma^{(1d)}$ can be described by the standard Drude formula 
\begin{align}
\sigma^{(1d)}=(2\pi R\nu^{(0)}_F)e^2D_0=e^2{\cal L}_{\rm loc},\qquad D_0=v_Fl/2,
	\label{main-cond1b}
\end{align}
where  $D_0$ is the diffusion coefficient, and $\nu_F^{(0)}=p_F/2\pi v_F$ is the density of states of an infinite two-dimensional metal at the Fermi surface (without ``spin two''). The factor $2\pi R\nu^{(0)}_F=(p_FR)/v_F=N_{\rm ch}/v_F$ is nothing else but the effective one-dimensional density of states in the wire. 

Certainly, there is no place  for the dependence on $\Phi$ in the result \eqref{main-cond1b}, such a dependence  can only show up in the corrections to the semiclassical approximation. There are three types of these corrections:

1. ``Thermodynamic'' corrections. Here we gather  all the oscillatory  effects, that arise already in the ideal tube without any impurities at all -- corrections to the density of states and similar. In the leading order in the semiclassical parameter $Q\gg 1$ the largest of these corrections are $\sim Q^{-1/2}\ll 1$, but we will need also the smaller ones. 

2. ``Kinetic'' corrections -- they arise due to oscillatory corrections to the single-impurity scattering amplitude. Usually the single-impurity effects only lead to an irrelevant renormalization of  parameters of an impurity. In our case, however, due to nontrivial topology of the system, these corrections are highly sensitive to the magnetic flux $\Phi$ and exhibit oscillatory behaviour. The relative amplitude of kinetic corrections (as well as the thermodynamic ones) does not depend on the concentration of impurities, its smallness comes mainly from the semiclassical parameter. The kinetic corrections are proportional to the scattering amplitude, they arise only beyond the Born approximation. In this Letter we will restrict our consideration to the first Born approximation, so that the kinetic corrections will not be taken into account in what follows.

3. Localization corrections. They stem from the effect of interference of scattering events at different impurities, their relative amplitude is proportional to $n^{(2)}_{\rm imp}$ and is small in parameter $1/(p_Fl)\ll 1$.  The localisation  corrections will also be neglected in this Letter.
	
	{\bf Conductivity in the Born approximation}. Consider randomly placed short range impurities with concentration $n^{(2)}_{\rm imp}$ (that is, the average number of impurities per unit area of the cylinders surface). Their potential is
\begin{align}
	\hat{H}_{\rm imp}=\sum_i V\delta({\bf r-r_i})
	\label{imp-potential}
\end{align}
where ${\bf r_i}$ ($z=z_i$, $\phi=\phi_i$) is a position of $i$-th impurity on the cylinder. In this paper we  consider only the case of weak impurities, with  small dimensionless strength
\begin{align}
	\lambda=\pi\nu_F^{(0)}V\ll 1
	\label{imp-potential1}
\end{align}
so that the Born approximation is valid.

For point impurities the scattering is isotropic and therefore the transport time coincides with the simple decay time. The latter for a state $(mk)$ can be written with the help of the Fermi golden rule
\begin{align}
	\tau_{k,m}^{-1}=2\pi	n_{\rm imp}^{(2)}\sum_{k'm'}|V_{kk'mm'}|^2\delta[E(km)-E(k'm')]
	\label{life-time}
\end{align}
Since for the potential \eqref{imp-potential} matrix elements $|V_{kk'mm'}|^2$ depends neither on $km$, nor on $k'm'$, we can write
\begin{align}
	\tau_{k,m}^{-1}=2\pi n^{(2)}_{\rm imp}V^2\nu_F
	\label{life-time1}
\end{align}
where 
\begin{align}
\nu_F	\equiv \sum_{mk}\delta(E_{mk}-E_F)=\nu_F^{(0)}(1+\Delta)	
,\label{green3rr}
\end{align}
is the density of states at the Fermi level for an ideal tube. Here  $\Delta$ is the correction, arising due to the finite radius of the cylinder.

It is important to note, that the decay rate in the leading approximation (i.e., without taking into account the correction $\Delta$) is isotropic: it does not depend on the angle $\theta$ between the momentum and the cylinder axis. 
	
	Within the Born approximation one can write for the conductivity
	\begin{align}
	\sigma=e^2(2\pi R)\sum_{mk}v_z^2(mk)\tau_{\rm tr}\delta[E(km)-E_F]=\nonumber\\=e^2\frac{\tau_{\rm tr}v_F^2}{2}(2\pi R)\tilde{\nu}_F
	\label{life-time2}
\end{align}
where the ``transport density of states'' $\tilde{\nu}_F$ is defined as
\begin{align}
	\tilde{\nu}_F=2\sum_{mk}(k/p_F)^2\delta\left(E_{mk}-E_{F}\right)=\nu_F^{(0)}(1+\tilde{\Delta})
\end{align}
where $v_{mk}=v_Fk/p_F$ is the velocity in the $z$ direction and $\tilde{\Delta}$ is, again,  the correction, arising due to the finite radius of the cylinder.

{\bf Thermodynamic corrections}. Corrections $\Delta$ and $\tilde{\Delta}$ are the only sources for the oscillations of the resistivity within the Born approximation.
Here we calculate them with the help of the Poisson summation formula.

An exact expression for $\nu_F$ reads
\begin{align}
	\nu_F=\sum_{mk}\delta\left(E_{F}-[k^2+(m+\Phi/2\Phi_0)^2/R^2]/2M\right)=\nonumber\\=\sum_{n=-\infty}^{\infty}\int\frac{d^2{\bf p}}{(2\pi)^2}\delta\left(E_{F}-p^2/2M\right)\nonumber\\\times\exp\left\{ 2\pi in(pR\sin\theta+\Phi/2\Phi_0)\right\},
\end{align}
where we have introduced effective mass $M=p_F/v_F$ and two-dimensional momentum: $p_z=k=p\cos\theta$ and $p_x=(m+\Phi/2\Phi_0)/R=p\sin\theta$, $\theta$ being the angle between the momentum and the cylinders' axis. Performing a trivial integration over $p$ and introducing parameter $Q\equiv \pi p_FR\gg 1$, we get
\begin{align}
	\nu_F=\nu_F^{(0)}\left\{1+2\sum_{n=1}^{\infty}\int_{-\pi}^{\pi}\frac{d\theta}{2\pi}\cos (2nQ\sin\theta)\cos (\pi n\Phi/\Phi_0)\right\}\nonumber
\end{align}
Thus
\begin{align}
	\Delta=2\sum_{n=1}^{\infty}J_0(2nQ)\cos (\pi n\Phi/\Phi_0)
\end{align}
where $J_0$ is the Bessel function. Using the two leading terms of its asymptotics, 
we get
\begin{align}
\Delta=\sum_{n=1}^{\infty}\left\{A(2nQ)\cos(2nQ-\pi/4)+\right.\nonumber\\+\left.B(2nQ)\sin(2nQ-\pi/4)\right\}\cos (\pi n\Phi/\Phi_0)
\end{align}
where
\begin{align}
A(x)=2\sqrt{\frac{2}{\pi x}},\quad B(x)=2\sqrt{\frac{2}{\pi x}}\frac{1}{8x}
\end{align}
In the leading approximation in $Q^{-1/2}$
\begin{align}
	\Delta=\sum_{n=1}^{\infty}\frac{2}{(\pi Qn)^{1/2}}\cos(2Q n-\pi/4)\cos(\pi n\Phi/\Phi_0)
	\label{delta0}
\end{align}	
but we will also need the subleading corrections in what follows.

For the transport density of states we have
\begin{align}
	\tilde{\nu}_F=2\sum_{mk}(k/p_F)^2\nonumber\\\times\delta\left(E_{F}-[k^2+(m+\Phi/2\Phi_0)^2/R^2]/2M\right)=\nonumber\\=\nu_F^{(0)}(1+\tilde{\Delta}),
\end{align}
where
\begin{align}
\tilde{\Delta}=2\sum_{n=1}^{\infty}\left\{J_0(2nQ)+J_2(2nQ)\right\}\cos (\pi n\Phi/\Phi_0).
\end{align}
Using the asymptotics of the Bessel functions, we can finally write
\begin{align}
\tilde{\Delta}=	
\sum_{n=1}^{\infty}\left\{\tilde{A}(2nQ)\cos(2nQ-\pi/4)+\nonumber\right.\\\left.+\tilde{B}(2nQ)\sin(2nQ-\pi/4)\right\}\cos (\pi n\Phi/\Phi_0),\\ \tilde{A}(x)=2\sqrt{\frac{2}{\pi x}}\frac{3}{4x^2}, \quad\tilde{B}(x)=2\sqrt{\frac{2}{\pi x}}\frac{2}{x}	
\end{align}
Thus, we conclude that the oscillation correction to $\tilde{\nu}_F$ is much smaller, than that for $\nu_F$: $\tilde{\Delta}\ll \Delta$. The reason is that the oscillations are provided by the states with very small $k$'s, whose velocity along $z$ is very small.

Now we are prepared to write a expression for the resistivity
\begin{align}
	\frac{1}{\rho}\approx \frac{1+\tilde{\Delta}}{1+\Delta}e^2D_0(2\pi R)\nu_F^{(0)},
	\label{life-time5}
\end{align}
\begin{align}
 \tau^{-1}_0=\frac{2\lambda^2}{\pi}\frac{n^{(2)}_{\rm imp}}{\nu_F^{(0)}},\qquad D_0=\frac{\tau_{0}v_F^2}{2},
	\label{life-time1}
\end{align}
are the perturbative decay rate and the diffusion coefficient of an infinite two-dimensional metal.

{\bf Corrections to the resistivity}. The relative oscillations of the resistivity  in the second order in $\tilde{\Delta}$ are 
\begin{align}
	\frac{\delta\rho}{\overline{\rho}}=\frac{\Delta-\tilde{\Delta}}{1+\tilde{\Delta}}=\Delta_{\rm tot}^{(1)}+\Delta_{\rm tot}^{(2)}+\cdots,
	\label{corr-tot}\\
\Delta_{\rm tot}^{(1)}=\Delta-\tilde{\Delta},\qquad	 \Delta_{\rm tot}^{(2)}=-(\Delta-\tilde{\Delta})\tilde{\Delta}
\label{corr1-corr2}
\end{align}
In the linear term the contribution $\Delta$ always dominates, so that $\Delta_{\rm tot}^{(1)}\approx\Delta$ is given by formula \eqref{delta0}.

In the second order correction $\Delta_{\rm tot}^{(2)}$ we should keep the ``diagonal'' terms of the form $\cos^2(2nQ-\pi/4)\cos^2n(\pi n\Phi/\Phi_0)$ or $\sin^2(2nQ-\pi/4)\cos^2(\pi n\Phi/\Phi_0)$. Only these terms contain contributions (equal to $(1/4)\cos (2\pi n\Phi/\Phi_0)$), that are insensitive to the fluctuations of the cylinder radius $R$, since they do not contain the factors, oscillating in $Q$:
\begin{align}
\Delta_{\rm tot}^{(2)}=-(\Delta-\tilde{\Delta})\tilde{\Delta}\approx\sum_{n=1}^{\infty}F(2nQ)\cos (2\pi n\Phi/\Phi_0)
	\label{life-time6x}
\end{align}
\begin{align}
F(x)=-\frac14\left\{(A-\tilde{A})\tilde{A}+(B-\tilde{B})\tilde{B}\right\}=\frac{6}{\pi x^3}
	\label{life-time6xk}
\end{align}
In a homogeneous tube the correction $\Delta_{\rm tot}^{(2)}$ should be neglected, compared to  $\Delta_{\rm tot}^{(1)}$, but, as we will see, it can become dominant, if fluctuations of radius are taken into account.

The series \eqref{life-time6x} converges for all $\Phi$, therefore the correction $\Delta_{\rm tot}^{(2)}$ is always relatively small due to the factor $Q^{-3}$. It is not the case of the correction $\Delta_{\rm tot}^{(1)}$, since  
the series \eqref{delta0} demonstrates a square-root divergence at certain values of $\Phi$. Below we discuss the behaviour of the resistivity close to the singularities and explain their physical origin.

{\bf Singularities in $\rho(\Phi)$}. It is convenient to rewrite the result \eqref{delta0} in the form
\begin{align}
	\Delta=(\pi Q)^{-1/2}\nonumber\\\times{\rm Re}\left\{S(2nQ+\pi \Phi/\Phi_0)+S(2nQ-\pi \Phi/\Phi_0)\right\}
	\label{life-time6ber}
\end{align}	
where
\begin{align}
	S(z)\equiv e^{-i\pi /4}\sum_{n=1}^{\infty}\frac{e^{inz}}{\sqrt{n}}
	\label{life-time6bei}
\end{align}	
This function has a singularity at $z\to 2\pi N$ for any integer $N$. In the vicinity of the singularity, where $\delta z\equiv z-2\pi N\ll 1$, one can write
\begin{align}
	S(z)=e^{-i\pi /4}\sum_{n=1}^{\infty}\frac{e^{in\delta z}}{\sqrt{n}}\approx e^{-i\pi /4}\int_0^{\infty}\frac{dn}{\sqrt{n}}e^{in\delta z} =\nonumber\\ =\frac{e^{-i\pi /4}}{\sqrt{-i\delta z}}\int_0^{\infty}\frac{e^{-t}dt}{\sqrt{t}}=\sqrt{\frac{\pi}{\delta z}}
	\label{life-time6beia}
\end{align}	
Thus, $S(z)$ is real and positive for $\delta z >0$ and it is purely imaginary for $\delta z <0$. Then
\begin{align}
\Delta\approx \left[2\pi Q(Q/\pi\pm\Phi/2\Phi_0-N)\right]^{-1/2}\nonumber\\\times\theta(Q/\pi\pm\Phi/2\Phi_0-N)
	\label{life-time6ber}
\end{align}	
 The physical reason for this square-root divergence is as follows. In our ideal quasi-one dimensional system we have a series of one-dimensional bands and, when the Fermi energy $E_F$ crosses the edge $E_N$ of one of these bands, the density of states diverges (as it should be in one dimension!):
\begin{align}
\nu_F=\sum_N\nu_N^{(1d)}(E_F),\nonumber\\	\nu^{(1d)}_N(E_F)\sim\left(\frac{\nu_F^{(0)}}{E_N-E_F}\right)^{1/2}\theta(E_N-E_F)
\end{align}
\begin{align}
	\Delta\approx \frac{1}{2\pi R\sqrt{2\pi \nu_F^{(0)}(E_N-E_F)}}\theta(E_N-E_F)
\end{align}
A natural question arises about a possible mechanism for smoothing the singularities in the $\rho(\Phi)$-dependence. The most obvious one is smearing of the singularities already in the density of states $\nu_F$. Such a smearing should occur, when the wave-length of an electron in a corresponding one-dimensional band becomes of order of the mean free path $l$. At low concentration of impurities, however, $l$ is very large and another mechanism of smoothing seems to be more effective -- the non-Born corrections to the scattering amplitude, which dramatically increase close to the singularities and lead to effective suppression of scattering probability, inspite of the divergence of the density of final states. This is a completely single-impurity effect, so it does not depend on the impurity concentration (and, therefore, on $l$) at all. 

All this singularity story is only actual for a homogeneous tube, however. As we will soon see, the fluctuations of the radius destroy the singularities anyway, and, therefore, in this Letter we will not discuss the singularity physics in detail and postpone it for a separate paper.

{\bf Fluctuations of the radius}. Now we consider a hollow metallic cylinder with fluctuating radius $R(z)$. For simplicity we will assume that the cylindrical symmetry is conserved, though this assumption is not essential: it is only the local circumference $c(z)$, that matters, so that one can define effective  local radius, as $R(z)\equiv c(z)/2\pi$. The fluctuations of $R(z)$ are taken in a gaussian form:
\begin{equation}
{\cal P}\{R(z)\}=\exp\left\{-\frac12\int dzdz'Q(z-z')\delta R(z)\delta R(z')\right\},\nonumber
\end{equation}
\begin{eqnarray}
\delta R(z)=R(z)-\overline{R},\quad\langle\delta R(z)\delta R(z')\rangle=K(z-z'),\nonumber\\ \hat{K}\equiv\hat{K}^{-1}=(\Delta R)^2f(|z-z'|/\xi),\nonumber
\end{eqnarray}
$f(x)$ being some function with $f(0)=1$ and characteristic scale $\sim 1$.
The fluctuations are supposed to be relatively small and smooth: 
\begin{equation}
\Delta R\ll \overline{R},\quad \xi\gg \overline{R},\label{ab7}
\end{equation}

So smooth fluctuations only lead to an exponentially (in parameter $\xi k_F$) small scattering of electrons and thus, do not give any considerable contribution to the decay rate. On the other hand, as we will see, they strongly suppress the Aharonov-Bohm oscillations.

For simplicity we discuss only the case of very long correlation length:
	$l\ll\xi\ll L$.
In this case one can think about our system as a sequence of short subwires  with lengths $L_i<\xi$, each of them having its own radius  $R_i$, randomly distributed according to the Gaussian law. Each subwire has its own resistance ${\cal R}_i=\rho(R_i)L_i$ and the total resistance ${\cal R}$ of the system of subwires, connected in series, is then
\begin{align}
	{\cal R}=\sum_i {\cal R}_i=\sum_i \rho(R_i)L_i=L\overline{\rho}, \\ \overline{\rho}=\int \frac{dR}{\sqrt{2\pi}\Delta R}\exp\left\{-\frac{(R-\overline{R})^2}{2\Delta R^2}\right\} \rho(R)dR
	\label{aver1}
\end{align}
Obviously, the averaging will dramatically suppress the Aharonov-Bohm oscillations, if the amplitude of fluctuations $\Delta R\gg p_F^{-1}$. Substituting \eqref{corr-tot}  \eqref{life-time6x} and \eqref{delta0} to \eqref{aver1}, and performing  trivial integration,  we arrive at the final result \eqref{life-time6u}.

{\bf Conclusion}. We have shown that the character of oscillations in the $\rho(\Phi)$ dependence for a cylinder with the fluctuating radius is controlled by the parameter $q=\pi\Delta R k_F$, where $\Delta R$ is the amplitude of fluctuations.

1. For $q\ll 1$ the oscillations are $2\Phi_0$-periodic, highly unharmonic, with sharp asymmetric maxima:
\begin{align}
\rho=\rho_0\left(1+\sum_{n=1}^\infty \frac{2}{ \sqrt{ \pi n\overline{Q}}}\cos (2n\overline{Q}+\pi/4)\cos(\pi n\Phi/\Phi_0)\right)
\nonumber
\end{align}
The relative amplitude of oscillations at maxima depends on details of the impurity potential and may be large.

2. For $q\sim 1$ the oscillations remain $2\Phi_0$-periodic, but become almost harmonic and have small amplitude $A_1\sim \overline{Q}^{-1/2}\ll 1$:
\begin{align}
\rho\approx\rho_0\left(1+\frac{2}{ \sqrt{ \pi \overline{Q}}}e^{-2q^2}\cos (2\overline{Q}+\pi/4)\cos(\pi \Phi/\Phi_0)\right)\nonumber
\end{align}

3. For $q\gg 1$ the oscillations become $\Phi_0$-periodic,  almost harmonic, and have still smaller amplitude $A_2\sim \overline{Q}^{-3}\ll A_1$:
\begin{align}
\rho\approx\rho_0\left(1+\frac{3}{2\pi\overline{Q}^{3}}\cos(2\pi \Phi/\Phi_0)\right)\nonumber
\end{align}

The $\Phi_0$-periodic oscillations are due to interference of  oscillating corrections to two different thermodynamic quantities: to the standard density of states and to the ``transport density of states''. Therefore the effect has additional semiclassical smallness (higher power of $\overline{Q}^{-1}$). It is important to stress, however, that, in contrast with the Altshuler, Aronov and Spivak effect, our one does not involve the interference between the scattering events at different impurities and, therefore, its relative amplitude does not contain small parameter $(p_Fl)^{-1}$ and does not depend on the concentration of impurities.

The author is indebted to M.~V.~Feigel'man, P.~A.~Ioselevich, and A.~Hutsalyuk for discussions. This work was supported by a Franco-Russian 
RFBR grant 13-02-91058-CNRS.

\end{document}